\documentclass[preprint,showpacs,12pt,preprintnumbers,amsmath,amssymb,nofootinbib]{revtex4}
\usepackage{bbm}
\usepackage{amsfonts}
\usepackage{booktabs}
\usepackage{mathrsfs}
\usepackage{epsfig}
\usepackage{graphicx}% Include figure files
\usepackage{dcolumn}% Align table columns on decimal point
\usepackage{bm}% bold math
\usepackage{amsmath}
\usepackage{slashed}
\usepackage{subfigure}

\let\jnfont=\rm
\def\NPB#1,{{\jnfont Nucl.\ Phys.\ B }{\bf #1},}
\def\PLB#1,{{\jnfont Phys.\ Lett.\ B }{\bf #1},}
\def\EPJC#1,{{\jnfont Eur.\ Phys.\ Jour.\ C }{\bf #1},}
\def\PRD#1,{{\jnfont Phys.\ Rev.\ D }{\bf #1},}
\def\PRL#1,{{\jnfont Phys.\ Rev.\ Lett.\ }{\bf #1},}
\def\MPLA#1,{{\jnfont Mod.\ Phys.\ Lett.\ A }{\bf #1},}
\def\JPG#1,{{\jnfont J.\ Phys.\ G}{\bf #1},}
\def\CTP#1,{{\jnfont Commun.\ Theor.\ Phys.\ }{\bf #1},}
\def\ZPC#1,{{\jnfont Z.\ Phys.\ C }{\bf #1},}
\def\JHEP#1,{{\jnfont JHEP \ }{\bf #1},}
\def\Rv{\not{\hbox{\kern-1pt $R$}}}
\def\p{\not{\hbox{\kern-3pt $p$}}}

\begin{document}

\title{Higgs couplings and Naturalness in the littlest Higgs model with T-parity at the LHC and TLEP}% Force line breaks with \\
%%\thanks{A footnote to the article title}%
\author{Bingfang Yang$^{1,2}$}
\author{Guofa Mi$^{1}$}
\author{Ning Liu$^{2}$}
\affiliation{$^1$ School of Materials Science and Engineering,
Henan Polytechnic University, Jiaozuo 454000, China\\
$^2$Institute of Theoretical Physics, Henan Normal University,
Xinxiang 453007, China
}%

\date{\today}% It is always \today, today,

\begin{abstract}
Motivated by the recent LHC Higgs data and null results in searches
for any new physics, we investigate the Higgs couplings and
naturalness in the littlest Higgs model with T-parity. By performing
the global fit of the latest Higgs data, electroweak precise
observables and $R_{b}$ measurements, we find that the scale $f$ can
be excluded up to 600 GeV at $2\sigma$ confidence level. The
expected Higgs coupling measurements at the future collider TLEP
will improve this lower limit to above 3 TeV. Besides, the top
parnter mass $m_{T_{+}}$ can be excluded up to 880 GeV at $2\sigma$
confidence level. The future HL-LHC can constrain this mass in the
region $m_{T_{+}} < 2.2$ TeV corresponding to the fine-tuning being
lager than 1\%.

\end{abstract}

\maketitle

%\tableofcontents

\section{\label{sec:level1}Introduction}

The discovery of a Higgs boson \cite{1} by the ATLAS\cite{2} and
CMS\cite{3} collaborations at the LHC marks a milestone of an effort
that has been ongoing for almost half a century and opens up a new
era of particle physics. The existing measurements \cite{4} and the
global fits to the ATLAS and CMS Higgs data within remarkable
precision\cite{5} agree with the standard model (SM) predictions.
This conclusion is consistent with the ATLAS and CMS null results in
searches for any new physics. However, the experiments of cold dark
matter\cite{6} and neutrino oscillations\cite{7} cannot be explained
in the framework of the SM so that they are supposed to provide
obvious evidence for the new physics beyond the SM. In particular,
the facts that the SM can be an effective theory valid all the way
up to the Planck scale and there is no symmetry protecting the
scalar masses lead to the naturalness problem, i.e., why the Higgs
boson mass is of the order of the electroweak scale and not driven
by the radiative corrections to the Planck scale, remains
unanswered.

Since the discovery of the Higgs boson the fine-tuning problem has
become even more intriguing. Among many new physics models, Little
Higgs models based on a collective symmetry breaking can provide a
natural explanation of the fine-tuning by constructing the Higgs as
a pseudo-goldstone boson. The littlest Higgs (LH) model\cite{LH} is
an economical approach to implement the idea of the little Higgs
theory. However, due to the large corrections to the electroweak
precision observables (EWPO) from the mixing of the SM gauge bosons
and the heavy gauge bosons, the original LH model is severely
constrained by precision electroweak data. This constraint can be
relaxed by introducing the discrete symmetry T-parity, which is
dubbed as littlest Higgs model with T-parity(LHT)\cite{LHT}.

With current data, all properties of the observed Higgs-like
particle turn out to be in rough agreement with expectations of the
SM\cite{10}, but there are still some rooms for the new physics
\cite{NP}, which may be ultimately examined at the LHC-Run2 and the
future Higgs factories \cite{H-factory}. Since top partner is
naturally related to the Higgs physics and plays an important role
in the naturalness problem, one can obtain constraints from the
Higgs data\cite{12}. In this work, we will discuss the Higgs
couplings and the naturalness problem in the LHT model at the LHC
and Triple-Large Electron-Positron Collider (TLEP)\cite{TLEP} by
performing a global fit of the latest Higgs data, $R_{b}$ and
oblique parameters, and give the current and future constraints to
the LHT parameters.

Recently, some similar works have been carried out in
Ref.\cite{range}. Different from these papers, we perform a
state-of-the-art global fit to obtain the indirect constraints on
the breaking scale and the top partner with a comprehensive way.
This method was widely used in the fit of the SM to the electroweak
precision data. So, it will be also meaningful to explore what might
happen in the LHT model with a global fit at future colliders. By
building an overall likelihood function for the constraints from the
EWPO, $R_{b}$ measurements and Higgs data, we can obtain a
well-defined statistical results of the exclusion limit on the
breaking scale. More importantly, we obtain the exclusion limit on
the top partner mass, which is obvious absent in other papers.

This paper is organized as follows. In Section II, we give a brief
description of the LHT model. In Section III, we present the
calculation methodology and the numerical results at the LHC and the
TLEP. Finally, we draw our conclusions in Section IV.

\section{ A brief review of the LHT model}
\noindent The LHT model is a non-linear $\sigma$ model based on the
coset space $SU(5)/SO(5)$, where the spontaneous symmetry breaking
is realised at the scale $f$ via the vacuum expectation value (VEV)
of an $SU(5)$ symmetric tensor $\Sigma$, given by
\begin{eqnarray}
\Sigma_0=\langle\Sigma\rangle
\begin{pmatrix}
{\bf 0}_{2\times2} & 0 & {\bf 1}_{2\times2} \\
                         0 & 1 &0 \\
                         {\bf 1}_{2\times2} & 0 & {\bf 0}_{2\times 2}
\end{pmatrix}.
\end{eqnarray}

The VEV of $\Sigma_0$ breaks the gauged subgroup $\left[ SU(2)
\times U(1) \right]^2$ of $SU(5)$ down to the SM electroweak
$SU(2)_L \times U(1)_Y$, which leads to new heavy gauge bosons
$W_{H}^{\pm},Z_{H},A_{H}$. After the EWSB, their masses up to
$\mathcal O(v^{2}/f^{2})$ are given by
\begin {equation}
M_{W_{H}}=M_{Z_{H}}=gf(1-\frac{v^{2}}{8f^{2}}),~~M_{A_{H}}=\frac{g'f}{\sqrt{5}}
(1-\frac{5v^{2}}{8f^{2}})
\end {equation}
with $g$ and $g'$ being the SM $SU(2)$ and $U(1)$ gauge couplings,
respectively. In order to match the SM prediction for the gauge
boson masses, the VEV $v$ needs to be redefined via the functional
form
\begin{equation}
v = \frac{f}{\sqrt{2}} \arccos{\left( 1 -
\frac{v_\textrm{SM}^2}{f^2} \right)} \simeq v_\textrm{SM} \left( 1 +
\frac{1}{12} \frac{v_\textrm{SM}^2}{f^2} \right) , \end{equation}
where $v_{SM}$ = 246 GeV is the SM Higgs VEV.

Under the unbroken $SU(2)_L \times U(1)_Y$ the Goldstone boson
matrix $\Pi$ is given by
\begin{equation}
\Pi = \left( \begin{array}{ccc} 0 & \frac{H}{\sqrt{2}} & \Phi
\\ \frac{H^\dagger}{\sqrt{2}}& 0 & \frac{H^T}{\sqrt{2}}\\ \Phi^\dagger
&\frac{H^*}{\sqrt{2}} & 0
\end{array}\right),
\end{equation}
where $H$ is the little Higgs doublet $(h^+,h)^T$ and $\Phi$ is a
complex triplet under $SU(2)_L$ which forms a symmetric tensor
\begin{equation}
    \Phi = \frac{- i}{\sqrt{2}} \begin{pmatrix} \sqrt{2} \phi^{++} & \phi^+ \\ \phi^+ & \phi^0 + i \, \phi^P \end{pmatrix} .
\end{equation}
$\phi^0$ and $\phi^P$ are both real scalars, whereas the $\phi^{++}$
and $\phi^+$ are complex scalars. The other Goldstone bosons are the
longitudinal modes of the heavy gauge bosons and therefore will not
appear in unitary gauge. The mass of $\Phi$ can be given by
\begin{equation}\label{tripletmass}
m_{\Phi}=\frac{2 m_H f}{v},
\end{equation}
where all components of the triplet are degenerate at the order we
are examining.

When T-parity is implemented in the quark sector of the model, we
require the existence of mirror partners with T-odd quantum number
for each SM quark. We denote the up and down-type mirror quarks by
$u_{H}^{i}$ and $d_{H}^{i}$, where $i$($i=1,2,3$) is the generation
index. After the EWSB, their masses up to $\mathcal O(v^{2}/f^{2})$
are given by
\begin{equation}
m_{d_{H}^{i}}=\sqrt{2}\kappa_if, ~~m_{u_{H}^{i}}=
m_{d_{H}^{i}}(1-\frac{\upsilon^2}{8f^2})
\end{equation}
where $\kappa_i$ are the diagonalized Yukawa couplings of the mirror
quarks. One can notice that the down-type mirror quarks have no
interactions with the Higgs.

In order to stabilize the Higgs mass, an additional T-even heavy
quark $T_{+}$ is introduced to cancel the large one-loop quadratic
divergences caused by the top quark. Meanwhile, the implementation
of T-parity requires a T-odd mirror partner $T_{-}$ with $T_{+}$.
The T-even quark $T_{+}$ mix with the SM top-quark and leads to a
modification of the top quark couplings relatively to the SM. The
mixing can be parameterized by dimensionless ratio
$R=\lambda_1/\lambda_2$, where $\lambda_1$ and $\lambda_2$ are two
dimensionless top quark Yukawa couplings. This mixing parameter can
also be used by $x_{L}$ with
\begin{equation}
x_{L}=\frac{R^{2}}{1+R^{2}}
\end{equation}
Considering only the largest corrections induced by EWSB, their
masses up to $\mathcal O(v^{2}/f^{2})$ are then given by
\begin{eqnarray}
&&m_t=\lambda_2 \sqrt{x_L }v \left[ 1 + \frac{v^2}{f^2} \left(
-\frac{1}{3} + \frac{1}{2} x_L \left( 1 - x_L \right) \right)
\right]\\
&&m_{T_{+}}=\frac{f}{v}\frac{m_{t}}{\sqrt{x_{L}(1-x_{L})}}\left[1+\frac{v^{2}}{f^{2}}\left(\frac{1}{3}-x_{L}(1-x_{L})\right)\right]\\
&&m_{T_{-}}=\frac{f}{v}\frac{m_{t}}{\sqrt{x_{L}}}\left[1+\frac{v^{2}}{f^{2}}\left(\frac{1}{3}-\frac{1}{2}x_{L}(1-x_{L})\right)\right]
\end{eqnarray}

The corrections to the Higgs couplings of the other two generations
of T-even (SM-like) up-type quarks up to $\mathcal{O} \left(
v_{SM}^4/f^4 \right)$ are given by
\begin{equation}
    \frac{g_{h \bar{u} u}}{g_{h \bar{u} u}^{SM}} = 1-\frac{3}{4}
        \frac{v_{SM}^{2}}{f^{2}}-\frac{5}{32} \frac{v_{SM}^{4}}{f^{4}}
        \qquad u \equiv u,c.
\end{equation}

For the T-even (SM-like) down-type quarks and charged leptons, the
Yukawa interaction have two possible constructions\cite{13}. The
corresponding corrections to the Higgs couplings up to $\mathcal{O}
\left( v_{SM}^4/f^4 \right)$ are given by ($d \equiv
d,s,b,l^{\pm}_i$)
\begin{eqnarray}
    \frac{g_{h \bar{d} d}}{g_{h \bar{d} d}^{SM}} &=& 1-
        \frac{1}{4} \frac{v_{SM}^{2}}{f^{2}} + \frac{7}{32}
        \frac{v_{SM}^{4}}{f^{4}} \qquad \text{Case A} \nonumber \\
    \frac{g_{h \bar{d} d}}{g_{h \bar{d} d}^{SM}} &=& 1-
        \frac{5}{4} \frac{v_{SM}^{2}}{f^{2}} - \frac{17}{32}
        \frac{v_{SM}^{4}}{f^{4}} \qquad \text{Case B}.
    \label{dcoupling}
\end{eqnarray}

The naturalness of the model can be quantified by the following
parameter ($\mu_{\text{obs}}^2$)\cite{14}:
\begin{equation} \label{eq:finetune}
    \Delta=\frac{|\delta \mu^2|}{\mu_{\text{obs}}^2}, \qquad \mu_{\text{obs}}^2= \frac{m_h^2}{2}.
\end{equation}
Here $m_h$ is the Higgs boson mass. In the LHT model, the dominant
negative log-divergent contribution to the Higgs mass squared
parameter comes from the top quark and its heavy partner $T_{+}$
loops\cite{14}
\begin{equation}
    \delta \mu^2 = -\frac{3 \lambda_t^2 m_{T_{+}}^2}{8 \pi^2} \log{\frac{\Lambda^2}{m_{T_{+}}^2}}
\end{equation}
where $\Lambda=4 \pi f$ is the UV cut-off of the model, $\lambda_t$
is the SM top Yukawa coupling.

\section{CALCULATIONS AND NUMERICAL RESULTS}

In our numerical calculations, we take the SM input parameters as
follows \cite{15}:
\begin{eqnarray*}
m_t &=& 173.5{\rm ~GeV},~~m_{W}=80.385 {\rm ~GeV}, ~~\alpha(m_Z)=
1/127.918,~~\sin^{2}\theta_W=0.231.
\end{eqnarray*}

Our global fit is based on the frequentist theory. For a set of
observables ${\cal O}_i (i=1...N)$, the experimental measurements
are assumed to be Gaussian distributed with the mean value ${\cal
O}^{exp}_i$ and error $\sigma^{exp}_i$. The $\chi^2$ can be defined
as $\chi^2 = \displaystyle{\sum_{i}^{N}}\frac{({\cal O}^{th}_i-{\cal
O}^{exp}_i)^2}{{\sigma_i}^2}$, where $\sigma_i$ is the total error
both experimental and theoretical. The likelihood
$\cal{L}\equiv$exp$[-\displaystyle{\sum} \chi_{i}^{2}]$ for a point
in the parameter space is calculated by using the $\chi^2$
statistics as a sum of individual contributions from the latest
experimental constraints. The confidence regions are evaluated with
the profile-likelihood method from tabulated values of $\delta\chi^2
\equiv -2 \ln({\cal L} / {\cal L}_{max})$. In three dimensions,
68.3\% confidence regions (corresponding to 1$\sigma$ range) are
given by $\delta\chi^2 =3.53$ and 95.0\% confidence regions
(corresponding to 2$\sigma$ range) are given by $\delta\chi^2 =
8.02$.

Under few assumptions involving mainly flavour independence in the
mirror fermion sector, the LHT model can be parametrised by only
three free parameters, i.e., the scale $f$, the ratio $R$ and the
Yukawa couplings of the mirror quarks $\kappa_{j}$. Considering the
recent constraint from the searches for the monojet, we require the
lower bound on the Yukawa couplings of the mirror quarks are
$\kappa_{j}\geq0.6$\cite{16}. We scan over these parameters within
the following ranges \cite{ewpc,range}
\begin{eqnarray*}
500{\rm GeV}\leq f\leq 2000{\rm GeV},~~0.1\leq R\leq 3.3,~~0.6\leq
\kappa_{j}\leq 3.
\end{eqnarray*}
where we assume the three generations $\kappa_{j}$ are degenerate.
The couplings of the UV operators are set as $c_{s} = c_{t} = 1$.
The likelihood function $\mathcal{L}$ is constructed from the
following constraints:

(1)EWPO: These oblique corrections can be described in terms of the
Peskin-Takeuchi S, T and U parameters\cite{18}. Firstly, the top
partner can contribute to the propagators of the electroweak gauge
bosons at one-loop level. In contrast to $T_{+}$, the T-odd top
partner $T_{-}$ does not contribute to S, T, U parameters since it
is an $SU(2)_{L}$ singlet which does not mix with the SM top quark.
Secondly, the T-odd mirror fermions give a contribution to the T
parameter at one-loop, which can have a noticeable effect on the
EWPO due to a large number (twelve) of doublets in the SM; Thirdly,
another important correction to both the S and T parameters follows
from the modified couplings of the Higgs boson to the SM gauge
bosons. Finally, other possible contributions arise from new
operators which parametrize the effects of the UV physics on weak
scale observables. All these different contributions to the oblique
parameters should be summed up. We calculate $\chi^{2}$ by using the
formulae in Refs.\cite{ewpc,19} and adopting the experimental values
of S, T and U in the Ref.\cite{15}.

(2) $R_{b}$. The branching ratio $R_{b}$ is very sensitive to the
new physics beyond the SM, the precision experimental value of
$R_{b}$ may give a severe constraint on the new physics. In the LHT
model, there are new fermions and new gauge bosons, which can
contribute to the $Zb\bar{b}$ coupling and give corrections to the
$R_{b}$ at one-loop level\cite{20}. The final combined result from
the LEP and SLD measurements show $R_{b} = 0.21629\pm
0.00066$\cite{15}, which is consistent with the SM prediction
$R_{b}^{SM} = 0.21578^{+0.0005} _{-0.0008}$.

(3) Higgs data. The experimental results are given in terms of
signal strengths $\mu(X; Y)$, which is defined as the ratio of the
observed rate for Higgs process $X\to h \to Y$ relative to the
prediction for the SM Higgs, $\mu(X;Y)\equiv\frac{\sigma(X)BR(h\to
Y)}{\sigma(X_{SM})BR(h_{SM}\to Y)}$. We confront the modified Higgs
interactions and the one-loop contribution of the new particles in
the LHT model with the available Higgs data. We calculate the
$\chi^{2}$ values by using the public package
\textsf{HiggsSignals-1.2.0}\cite{21}, which includes 81 channels
from the LHC and Tevatron and these experimental data are listed in
Ref.\cite{data}. In our calculations, the Higgs mass $m_{h}$ is
fixed as 126 GeV. Note that for the Higgs data, the
\textsf{HiggsSignals} has provided the calculation of $\chi^2$,
where both experimental (systematic and statistical) uncertainties
as well as SM theory uncertainties are included.

%%%fig1.eps%%%%%%%%%%%%%%%%%%%%
\begin{figure}[h]
\centering
\includegraphics[width=3.2in,height=2.7in]{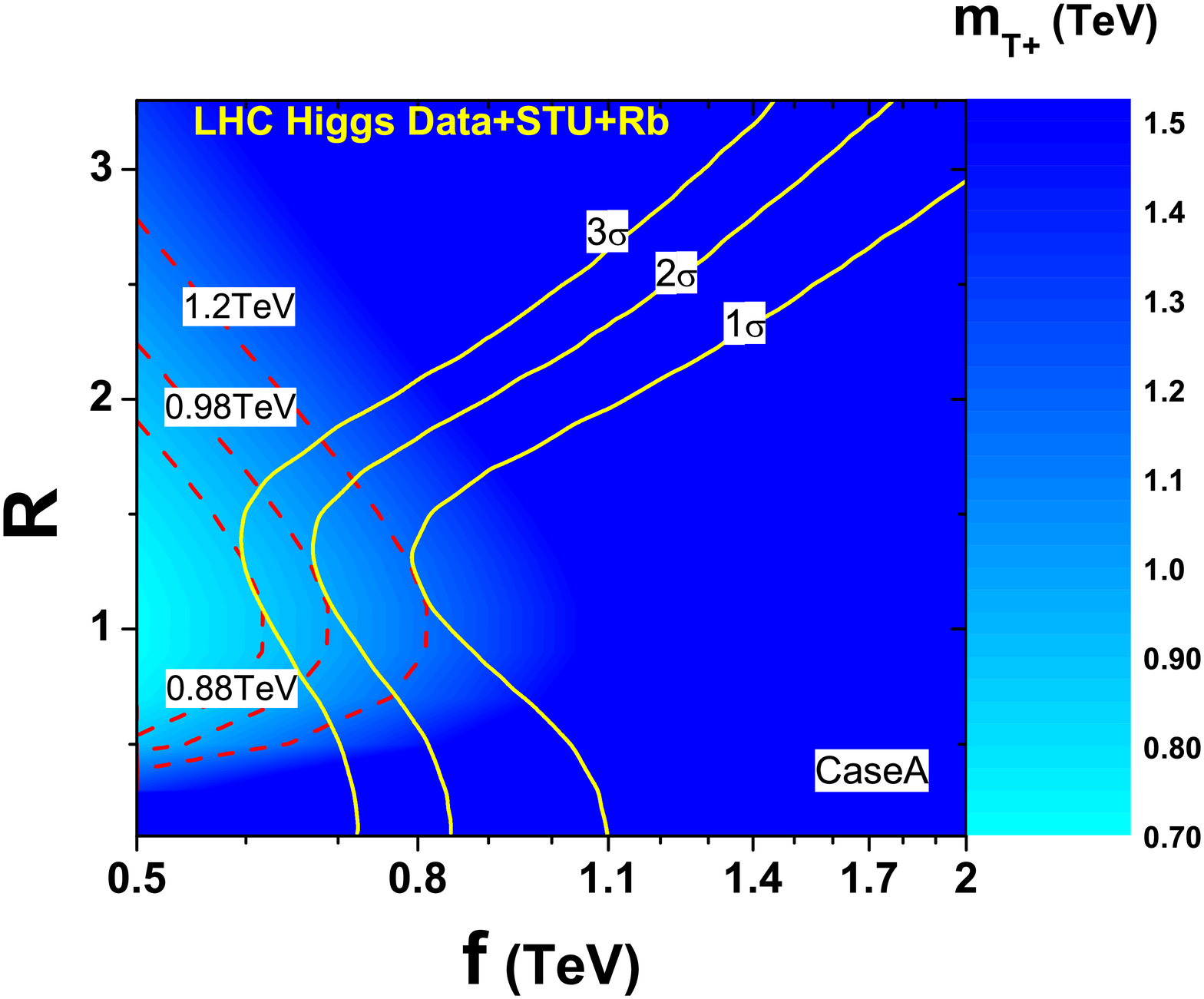}
\includegraphics[width=3.2in,height=2.7in]{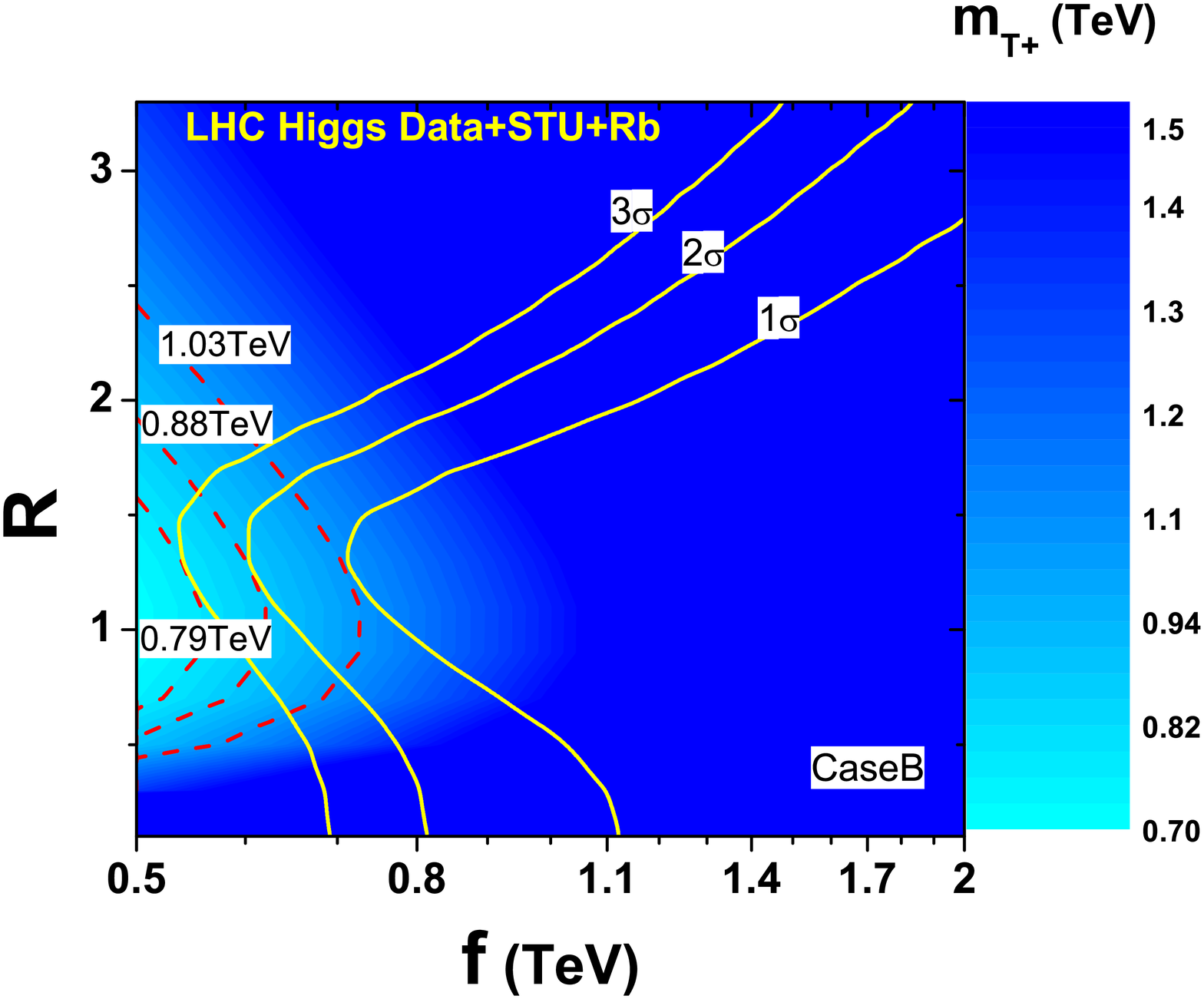}\vspace{-1cm}
\vspace{-0.3cm} \caption{The global fit of the constraints on the
LHT model in the $R-f$ plane for Case A and Case B. The yellow lines
from right to left respectively correspond to $1\sigma$, $2\sigma$
and $3\sigma$ exclusion limits.} \label{amt}
\end{figure}
%%%%%%%%%%%%%%%%%%%%%%%%%%%%%%%%
In Fig.1, we show the results of the global fit to the above three
kinds of constraints in the plane of $R$ versus $f$ for Case A and
Case B, respectively. We can see that the lower bound on the
symmetry breaking scale at 95\% C.L. is
\begin{eqnarray}
&&f>670\rm GeV \qquad \text{Case A}, \\
&&f>600\rm GeV \qquad \text{Case B}.
\end{eqnarray}
The constraints are stronger than the electroweak precision
constraints in Ref.\cite{ewpc}, which is because the main constraint
here comes from the Higgs data. For the top partner mass, we can see
that the combined indirect constraints can exclude $m_{T_{+}}$ at
95\% C.L. up to
\begin{eqnarray}
&&m_{T_{+}}>980\rm GeV \qquad \text{Case A}, \\
&&m_{T_{+}}>880\rm GeV \qquad \text{Case B}.
\end{eqnarray}
It's worth noting that they are stronger than the lower bound set by
the ATLAS direct searches for the $SU(2)$ singlet top partner,
$m_{T} > 640$ GeV \cite{23}. Our study may play a complementary role
to the direct searches in probing top partner.
\begin{table}[htbp]
\caption{Expected precision on the Higgs couplings to quarks and
vector bosons at the HL-LHC and the TLEP. }
\begin{center}
\begin{tabular}{|c|c|c|}\hline
Facility      &HL-LHC & TLEP    \\
\hline
$\sqrt{s}$     &14TeV & 240GeV      \\
$\int{\cal L}dt$  &3000(fb$^{-1}$) &  10000(fb$^{-1}$)  \\
\hline
 $\kappa_{\gamma}$  &$2-5\%$ & 1.7\%  \\
 $\kappa_{g}$     &$3-5\%$ & 1.1\%  \\
 $\kappa_{W}$     &$2-5\%$ & 0.85\%  \\
 $\kappa_{Z}$     &$2-4\%$ & 0.16\% \\\hline
 $\kappa_{u}$       &$7-10\%$ & $-$\\
 $\kappa_{d}$       &$4-7\%$ & $-$\\
 $\kappa_{c}$     &$7-10\%$ & 1.0\%  \\
 $\kappa_{s}$     &$4-7\%$ & $-$\\
 $\kappa_{t}$     &$7-10\%$ & $-$   \\
 $\kappa_{b}$     &$4-7\%$ & 0.88\% \\
\hline
\end{tabular}
 \end{center}\label{tab:precision}
\end{table}

The expected precision for the Large Hadron Collider High-Luminosity
Upgrade (HL-LHC) and the TLEP are assumed in
Table-\ref{tab:precision}, which comes from the Table-14 and
Table-16 of the Higgs working group report\cite{24}.

%%%fig2.eps%%%%%%%%%%%%%%%%%%%%
\begin{figure}[h]
\centering
\includegraphics[width=6in,height=5in]{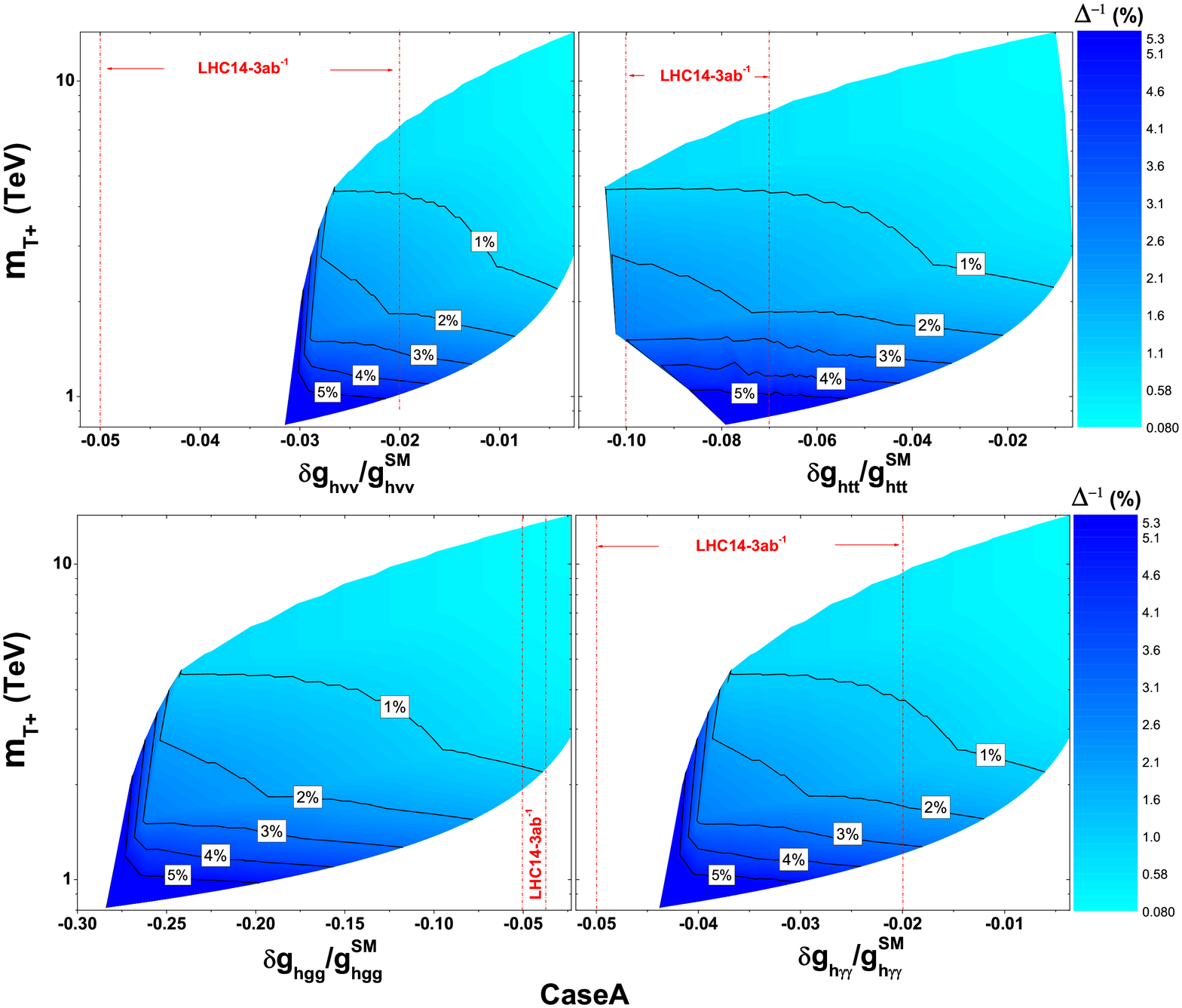}
\vspace{-0.3cm} \caption{The shifts of the Higgs couplings for the
samples in the $2\sigma$ allowed range in Fig.1 for Case A. The red
dash-dot lines represent the expected measurement uncertainties at
HL-LHC.} \label{couplinga}
\end{figure}
\begin{figure}[h]
\centering
\includegraphics[width=6in,height=5in]{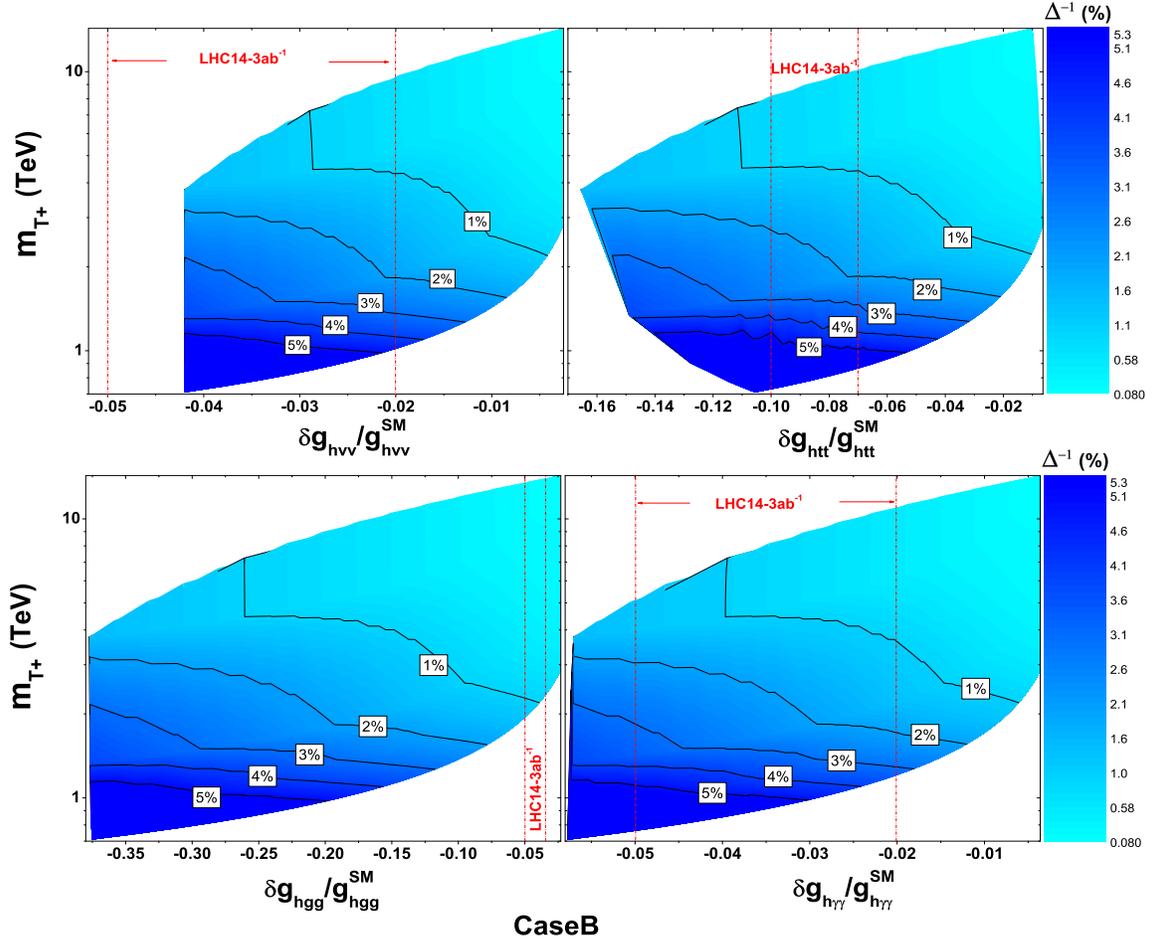}
\vspace{-0.3cm} \caption{The shifts of the Higgs couplings for the
samples in the $2\sigma$ allowed range in Fig.1 for Case B. The red
dash-dot lines represent the expected measurement uncertainties at
HL-LHC.} \label{couplingb}
\end{figure}

%%%%%%%%%%%%%%%%%%%%%%%%%%%%%%%%
In the LHT model, the loop-induced couplings $hgg$ and
$h\gamma\gamma$ can receive contributions from both the modified
couplings and the new particles. The decay $h\to gg$ can be
corrected by the modified $ht\bar{t}$ coupling and the loops of top
partner $T_{+}$ and T-odd mirror quarks. In addition to these
corrections involved in the decay $h\to gg$, the decay $h\to
\gamma\gamma$ can be also corrected by the modified $hWW$ coupling
and the loops of $W_{H}$, $\phi^{+}$, $\phi^{++}$ . Besides, the
couplings $hc\bar{c}$, $hs\bar{s}$, $hb\bar{b}$, $hZZ$ are also
modified, they can exert an effect on our fit.

In Fig.2 and Fig.3, we show the shifts of the Higgs couplings $hV
V$, $ht\bar{t}$, $hgg$, $h\gamma\gamma$ for the above samples in the
2$\sigma$ range. In order to investigate the observability, we
compare them with the corresponding expected measurement
uncertainties of the Higgs couplings in Table-\ref{tab:precision} at
HL-LHC with a luminosity of 3000 fb$^{-1}$. The value of the
fine-tuning for each point is also calculated by using the
Eq.(\ref{eq:finetune}). From Fig.2 and Fig.3, we can have some
observations as follows:

(1) The values of the fine-tuning for the samples are cornered to be
smaller than about 6\% by the above global fit;

(2) For the Higgs couplings $hVV$ and $ht\bar{t}$, they are
suppressed by the high order factor $\mathcal{O} \left( v^2/f^2
\right)$. The deviation of the Higgs couplings $g_{hVV}$ from the SM
predictions are at percent level and the deviation of the Higgs
coupling $g_{ht\bar{t}}$ from the SM prediction can reach over 10\%.

For the loop-induced couplings $g_{hgg}$ and $g_{h\gamma\gamma}$, on
one hand they are corrected by the high order factor, on the other
hand they are corrected by the loop contributions of the new
particles. For the effects of these loop diagrams, there are
cancelation between $t(W_{L})$ and the corresponding partner
$T_{+}(W_{H})$ so that the effective $g_{hgg}$ and
$g_{h\gamma\gamma}$ couplings are reduced. The deviation of the
Higgs coupling $g_{h\gamma\gamma}$ from the SM prediction is at
percent level, that is because the dominant contribution to the
coupling $g_{h\gamma\gamma}$ comes from the $W_{L}(W_{H})$ over the
$t(T_{+})$. The Higgs coupling $g_{hgg}$ from the SM prediction can
reach about 30\%, that is because the dominant contribution to the
coupling $g_{hgg}$ comes from the $ht\bar{t}$ coupling and
$t(T_{+})$ loops, where the contribution of $ht\bar{t}$ coupling
accounts for about 10\% and the contributions of $t(T_{+})$ loops
account for about 20\%. Furthermore, we can see that the deviations
for Case A are less than that for Case B, which originate from the
stronger suppression for the down-type fermion couplings to the
Higgs boson in Case B.

Furthermore, we can see that all changes of the Higgs couplings are
negative. In the LHT model, in order to cancel the quadratic
divergence of the Higgs mass, the heavy gauge bosons and the
additional heavy quark $T_{+}$ are introduced. This leads to
negative modification of the relevant couplings with respect to the
SM. Besides, the non-linear expansion of the model field suppresses
these couplings at the order ${\cal O}(v^{2}/f^{2})$;

(3)In Fig.2 and Fig.3, we attempt to show the expected constraints
from the future individual Higgs coupling meaurements on the top
partner and naturalness at the HL-LHC. The couplings $hVV$ and
$ht\bar{t}$ are modified at the order ${\cal O}(v^{2}/f^{2})$, which
can determine the scale $f$ and help us understand the nature of the
Higgs boson in the LHT model. Apart from this, the coupling $hgg$
can provide the information for the cancelation between $t$ and the
corresponding partner $T_{+}$, while the coupling $h\gamma\gamma$
can provide the information for the cancelation between $W_{L}$ and
the corresponding partner $W_{H}$. So, we can see that the
individual Higgs coupling meaurements can help us understand the
different parts of the LHT model.

(4)The future measurements of the $g_{hgg}$ coupling at the HL-LHC
will be able to exclude the $m_{T_{+}} < 2.2$ TeV, which corresponds
to the fine-tuning being lager than about 1\%. However, other
expected measurements, such as $g_{hVV}$, $g_{ht\bar{t}}$ and
$g_{h\gamma\gamma}$ couplings, can only improve the limits for the
top partner mass mildly.

%%%fig4.eps%%%%%%%%%%%%%%%%%%%%
\begin{figure}[h]
\centering
\includegraphics[width=3.2in,height=2.8in]{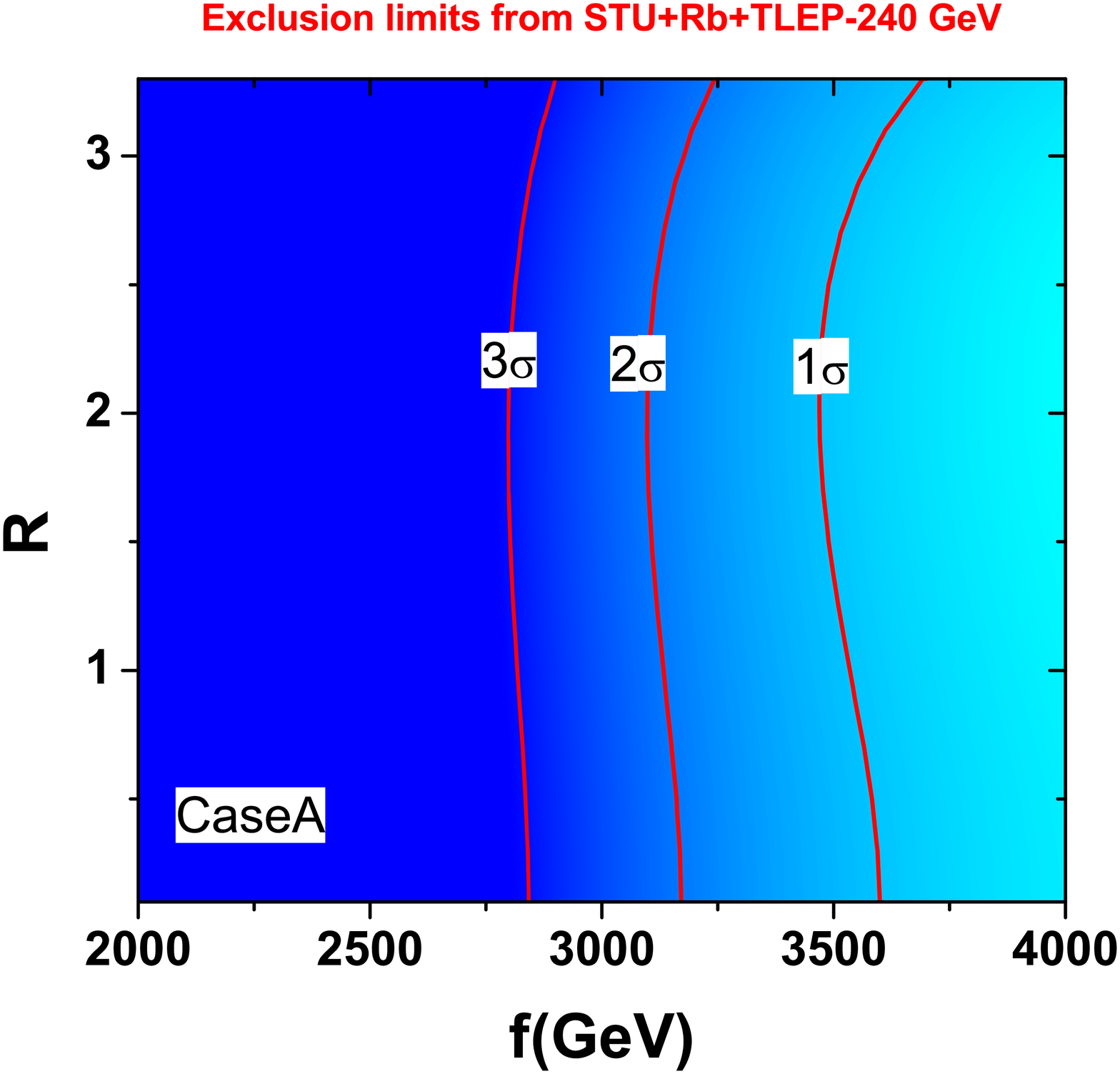}
\includegraphics[width=3.2in,height=2.8in]{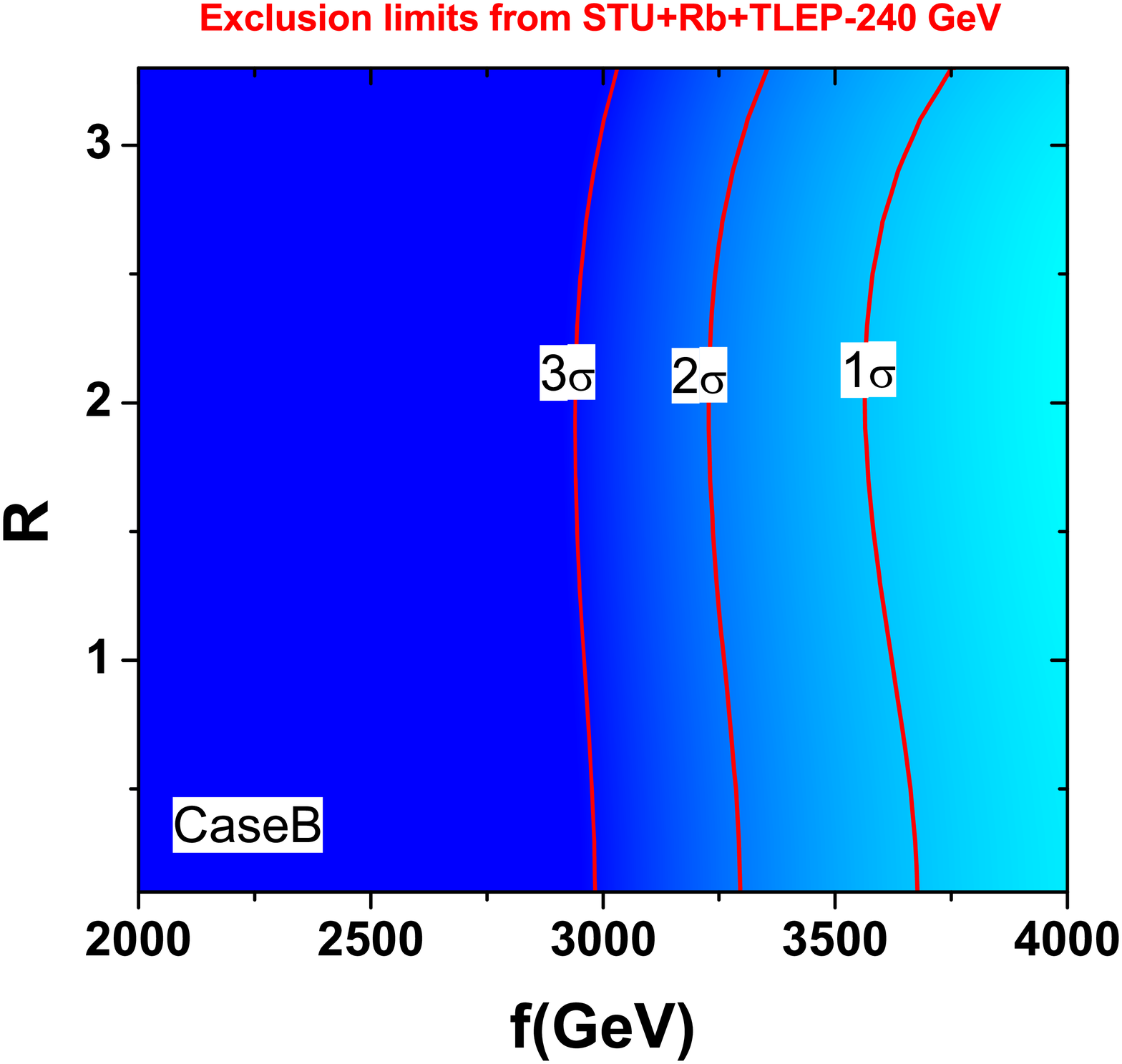}\vspace{-0.5cm}
\caption{The expected exclusion limits on the $R-f$ plane for Case A
and Case B from the global fit of EWPO, $R_{b}$ and TLEP.}
\label{tlep}
\end{figure}
%%%%%%%%%%%%%%%%%%%%%%%%%%%%%%%

In Fig.4, we present the prospect of improving the constraints on
the scale $f$ at a possible future Higgs factory TLEP with $\sqrt{s}
= 240$ GeV. In our fit, the $\chi^2$ can be defined as
\begin{eqnarray}
\chi^2 = \displaystyle{\sum_{i}^{N}}\frac{(\mu_i-1)^2}{{\sigma_i}^2}
\end{eqnarray}
where $\mu_{i}$ represents the signal strength prediction from the
LHT model and $\sigma_{i}$ represents the $1\sigma$ uncertainty i.e.
the expected measurement precision at the TLEP. We use the Snowmass
Higgs working group results to simply estimate the exclusion limits.
Given that the super-high luminosity of 10000 fb$^{-1}$ can be
achieved at the TLEP, we assume that all the measured Higgs
couplings will be the same as the SM predictions with the expected
measurement uncertainties in Table-\ref{tab:precision}. From the
Fig.4, we can see that the lower bound on the scale $f$ will be
pushed up to 3.1 TeV for Case A and 3.25 TeV for Case B at 95\%
C.L..

\section{\label{sec:level1}conclusions}

In this paper, we investigated the Higgs couplings and naturalness
in the LHT model under the available constraints from the current
Higgs data and the EWPO. By performing the global fit, we find that
the scale $f$ can be excluded up to 670 GeV for Case A and 600 GeV
for Case B at 2$\sigma$ level. The precise measurements of the Higgs
couplings at the future collider TLEP will constrain this limit to
above 3 TeV. Besides, the top partner mass $m_{T_{+}}$ can be
excluded up to 980 GeV for Case A and 880 GeV for Case B at
2$\sigma$ level. This mass can be constrained in the region
$m_{T_{+}} < 2.2$ TeV at the HL-LHC corresponding to the fine-tuning
being lager than 1\%.

\section*{Acknowledgement}
We would like to thank Marco Tonini for valuable discussions and
also thank Lei Wu for helpful suggestions. This work was supported
by the National Natural Science Foundation of China (NNSFC) under
grants Nos.11275245, 11305049, 11347140 and 11405047, by Specialized
Research Fund for the Doctoral Program of Higher Education under
Grant No.20134104120002, the Startup Foundation for Doctors of Henan
Normal University under contract No.11112 and the China Postdoctoral
Science Foundation under Grant No.2014M561987.

\end{document}